\documentclass[12pt,preprint]{aastex}%
\usepackage{natbib}
\usepackage{amsmath}
\usepackage{graphicx}%
\usepackage{amsfonts}%
\usepackage{amssymb}

\shorttitle{thin disks?}
\shortauthors{Meirelles et al.}

\begin{document}

\title{Are There Thin Accretion Disks?}

\author{Cesar Meirelles Filho}

\affil{Instituto de Astronomia, Geof\'{\i}sica e de Ci\^{e}ncias Atmosf\'{e}ricas \\
Universidade de S\~ao Paulo \\R. do Mat\~ao, 1226, 05508-090 S\~ao Paulo, SP, Brasil}

\email{cmeirelf@astro.iag.usp.br}

\author{Celso L. Lima, and Hideaki Miyake}

\affil{Instituto de F\'{\i}sica \\Universidade de S\~ao Paulo \\CP 66318, 05315-970 S\~ao Paulo, SP, Brasil}

\begin{abstract}
Total and internal energy requirements set constraints on the allowable
temperature on the disk. In a gas pressure dominated disk, they lead to a
cooling rate that decreases with temperature. These findings are used to show
that the flow is mildly supersonic, with the azimuthal Mach number moderately
exceeding one in a large extent of the disk. Only in the narrow outermost
region, of about $12\,\%$ of the disk extent, the Mach number greatly exceeds
one. It is also shown that, under most favorable conditions, radiation can ,
at the most, transport $20\,\%$ of the heat produced, close to the inner
radius. Far away, under the same conditions, this efficiency can rise to
$100\,\%$. This result depends only on boundary conditions, being independent
of the accretion rate and mass of the central object.The geometrical thin disk
approximation is violated everywhere in the disk, except in the narrow region
where, under the unlikely assumption of thermodynamical equilibrium or, at
least, local equilibrium, the flow is indeed supersonic.

\end{abstract}


\section{Introduction}

In the last 30 years, accretion disks have played a major role in theoretical
astrophysics. Part of this is due to the growing evidence of their presence in
almost every astrophysical environment. However, it would be unfair do not
recognize that this popularity is also due to physical picture of an accretion
disk set forward by \citet{ss73} ${\alpha}$-standard model. The simplicity of
the arguments that have led to the construction of this model is quite
striking, such that, despite some drawbacks after the long time in the road,
some authors keep using it as their main tool to explain data.

However, the model having thinness and hydrostatic equilibrium as one of the
main assumptions in its foundations, was meant to apply under very specific
conditions. These certainly would exclude disks with very high accretion
rates, or radiative pressure dominated disks. As a matter of fact, in that
situation the disk would be thermal and secularly unstable \citep{lig74}.
According to the picture given by the standard model, the region exposed to
swelling or instability would be the innermost one, which would exist whenever
the radiation pressure exceeds the gas pressure. In other words, for high
accretion rates. Another difficulty of the model is its impossibility of
reproducing the correct transonic nature of the flow close to the horizon
\citep{cha90}. During quite a long time these two issues have been a major
concern of the research in the area. Among models aiming to develop
alternative accretion disk models, it is worth mentioning \citet{pa80} model
which replaced the usual assumption of local energy balance by a global energy
requirement, under the assumption of critical radiative flux. Unquestionably,
one of the main contributions of this work was the pioneering introduction of
a pseudo Newtonian potential which simulates relativistic effects without any
resource to a burdensome mathematics. At that time, the development of a
theory of geometrically thick accretion disks around black holes has been
undertaken by \citet{ab78,ko78,ja80,be82}. Progress, concerning the
transonic nature of the flow, came in a much slower way through the
contributions from \citet{ko78,li80,pa81,lo82,mu83,ab89}. Since the flow has
to be continuous, the solution that incorporates the transonic nature of the
flow, close to the horizon, should match the solution far away, where,
supposedly, the disk is geometrically thin and Keplerian. However, this
matching introduces some arbitrariness, such as matching radius, and the
matching of the radial and angular velocities. The problem here is the angular
momentum distribution, which, close to the horizon, is highly dependent upon
the matching radius \citep{cha96}. Once this radius is determined, the
angular momentum follows practically constant inside the horizon, a fact which
suggests null boundary condition assumption for the torque \citep{mu83}.
Besides, if one adopts a more realistic model of disk, allowing for
dissipation, the situation may become much more complex. This is due to the
topology of the solutions near the critical and subcritical points, both of
them occurring in dissipative flows. Critical points satisfy regularity
conditions in the form of algebraic constraints, while for the subcritical
points they are given by integral requirements. In determining the character
of the solution near these points, the role of the viscosity is quite
decisive, such that there is a critical value of $\alpha$ above which no
stationary accretion is possible \citep{mu83}, a question contested by
\citet{ma84} and \citet{ab89}. Clearly, the matching of an ``inner disk''
with an ``outer disk'' is highly dependent on what kind of disk one assumes as
an outer disk, that is to say, geometrically thin or not, a fact that may be
seen in the works of \citet{li80} and \citet{lo82}.

Recently, these issues have been revived, connected with the question of ``no
torque'' inner boundary condition \citep{kr99, ga99,ag00,ap03}. As argued by
\citet{ap03}, the no torque inner boundary condition applies whenever the
disk thickness is small everywhere. However, results based on semianalytical
and numerical work employing advective dominated accretion flows (ADAF) and
convective dominated accretion flows (CDAF) show that accretion disks are
thick for very high and very low accretion rates \citep{ba01}. Thinness
would occur only for moderate accretion rates $(0.01\leq{L/L_{Edd}}\leq0.1)$.
It seems that, in order to bypass the issue of thermal balance, one adopts the
simplifying assumption of disk thinness. This is not justified a priori,
because even numerical simulations yield relatively thick outer boundary, not
far away from the inner one \citep{ap03}. However, these simulations suffer
from limitations, one of them being not including cooling processes. In the
literature, alledgedly, the assumption of highly supersonic character of the
Keplerian flow is used as the argument for the disk thinness. This is, for
sure, intuitive, but, by no means, unquestionable. Somehow, it seems that
thinness is favored by the usual assumption of local equality between heat
production and cooling by radiation, which overlooks the problem of energy
conservation in the disk. One of the advantages of doing so is to render the
problem algebraic, rather than differential , non local. In the solutions to
the disk equations, the ${\alpha}$ parameter doesn't enter with a high power
\citep{fran92}, which due to its inherent uncertainty, is very interesting.
At this point, we realize how decisive the use of the geometrical thinness, as
a boundary condition, is in determining the structure of the whole disk. If
cooling of the disk is not efficient, most of the viscously produced heat is
stored as proton internal energy and advected with the flow. The temperature
tends to the virial value and the height scale of the disk will be comparable
to the radial distance. More important, the internal energy of the plasma
approaches the rest energy and, at the inner radius, the sound velocity is
much greater than the radial velocity, being a substantial fraction of the
light velocity. Since the gas only cools after passing through the sonic point
\citep{mu86}, the sonic velocity keeps growing till the sonic point which,
in that situation, will be closer to the hole. Though the picture we have
given may be incomplete, it seems that, somehow, the importance of the disk
being thin or not has been overlooked. To justify this very pertinent point,
we would like to stick on the conclusions by \citet{ap03} and on the
possibility, cast by them, of non existence of thin geometrical disks in
nature. We, therefore, propose to study, in a more detailed way, the
geometrical thin disk assumption. Our procedure will be to use, from the very
beginning, energy requirements constraining both the temperature, the height
scale of the disk, viscous heating, and cooling of the disk. This, certainly,
is a global requirement that avoids the usual assumption of local equality
between viscous heat production and cooling. Since the picture of geometrical
thin disk is so profoundly rooted in our minds, in this work we shall focus
our attention only on that.

The plane of the paper is as follows. Section 2 presents the system we are
considering, the assumptions and the disk equations, including an outline of
the ingredients to relate angular momentum transport and dissipation to the
density and radial velocity. In section 3, the total and internal energy
equations are handled in such a way as to lead to an equation to the
temperature of the disk. In section 4, some approximations are made in order
to solve the equation for the temperature in conditions similar to those where
the standard model is supposed to apply; the main differences between our
solution and the standard one are emphasized. In section 5, general
considerations about the radiative transfer problem link the results to
specific radiative mechanisms. In section 6, the flow regimes that may emerge
are discussed in the context of different boundary conditions. In section 7,
the main conclusions are drawn.

\section{The System And The Disk Equations}

The system we consider is a disk around a primary compact object (a neutron
star, a black hole), of mass $M$. Self gravitation in the disk is negligible.
Working in a reference frame tied to the center of mass of the compact object,
a point on the disk is at a distance $R$ from the center and has coordinates
($x$, $y$, $z$) or ($r$, $\phi$, $z$). Of course, $R=\sqrt{\left(  r^{2}%
+z^{2}\right)  }$.

For the moment, we shall use Newtonian gravitation, and the potential at the
point ($r$, $\phi$, $z$) is%

\[
\Psi=-\frac{GM}{\sqrt{\left(  r^{2}+z^{2}\right)  }},
\]
G being the gravitational constant.

For the ${\phi}$ component of the velocity we will assume a Keplerian profile.
Therefore, we shall write%

\[
V_{\phi}=r\,\Omega_{k}\,,
\]
where ${\Omega}_{k}$ is the Keplerian angular velocity,%

\[
\Omega_{k}=\sqrt{\frac{GM}{r^{3}}}.
\]

Since we are assuming the disk in hydrostatic equilibrium,%

\begin{equation}
\frac{dP}{dz}=-\rho\,\Omega_{k}^{2}\,z, \label{eq1}%
\end{equation}
where ${\rho}$ is the density and $P$ is the pressure. Now, writing%

\[
P=P_{g}+P_{r},
\]
where%

\[
P_{g}=\rho\,\frac{k_{B}}{m_{H}}\,T
\]
is the gas pressure, $T$ is the temperature, and%

\[
P_{r}=\frac{F_{r}}{c}\frac{{\rho}\,{\sigma}_{T}\,{\ell}}{m_{H}}%
\]
is the radiative pressure. $F_{r}$ is the radiative flux, $\ell$ is the
semi-height scale of the disk, $k_{B}$ is the Boltzmann constant, $m_{H}$ is
the Hydrogen mass, $c$ is the velocity of light, and ${\sigma}_{T}$ is the
Thompson cross section for electron scattering. Assuming constant density
along $z$, and taking the average over $z$ in equation (\ref{eq1}), yields%

\begin{equation}
\frac{{{\Omega}_{k}}^{2}\,{\ell}^{2}}{3}=\frac{k_{B}}{m_{H}}\,T+\frac{F_{r}%
}{c}\frac{{\sigma}_{T}\,{\ell}}{m_{H}}\,. \label{eq2}%
\end{equation}

The continuity equation is written as%

\begin{equation}
\dot{M}=-4\,\pi\,r\,\ell\,V_{r}\,\rho\,, \label{eq3}%
\end{equation}
where ${\dot{M}}$ is the accretion rate, and $V_{r}$ is the radial velocity.

Now, we need to work out some relations between angular momentum transport and
dissipation to obtain the usual expressions for the density and radial
velocity. If $\dot{J}_{in}$, $\dot{J}_{0}$ and $\dot{J}_{out}$ are,
respectively, the inwards angular momentum transport (per unit time), the rate
at which the angular momentum flows into the central compact object and the
outwards flux of angular momentum, we have%

\begin{equation}
\dot{J}_{in}=\dot{J}_{0}+\dot{J}_{out}, \label{eq4}%
\end{equation}
and we may write%

\begin{equation}
\dot{J}_{in}=\dot{M}\,\sqrt{G\,M\,r}, \label{eq5}%
\end{equation}%

\begin{equation}
\dot{J}_{0}=\dot{M}\,\sqrt{G\,M\,r_{1}} \label{eq6}%
\end{equation}
and%

\begin{equation}
\dot{J}_{out}=2\,\pi\,r\,\ell\,\tau_{r\,\phi}, \label{eq7}%
\end{equation}
where ${\tau}_{r\,\phi}$ is the stress tensor given by%

\begin{equation}
\tau_{r\,\phi}=-2\,\eta\,\sigma_{r\,,\phi}, \label{eq8}%
\end{equation}
${\sigma}_{r\,,\phi}$ being the rate of strain tensor, given by%

\begin{equation}
\sigma_{r\,,\phi}=-\frac{r}{2}\,\frac{\partial{{\Omega}_{k}}}{{\partial r}}.
\label{eq9}%
\end{equation}

Then, the angular momentum conservation may be written as%

\begin{equation}
\dot{M}\,\sqrt{G\,M\,r}=\dot{M}\,\sqrt{G\,M\,r_{1}}+2\,\pi\,r\,\ell
\,\tau_{r\,\phi}. \label{eq10}%
\end{equation}

In the above equation, we have assumed null boundary condition for the torque
at the inner radius $r= r_{1}$. This yields for the energy dissipation,%

\begin{equation}
q^{+}=\frac{3}{{8\,{\pi}}}\,\frac{{G\,M\,{\dot{M}}}}{r^{3}}\,\left(  1-\left(
\frac{r_{1}}{r}\right)  ^{1/2}\right)  , \label{eq11}%
\end{equation}
and for the density%

\begin{equation}
\rho=\frac{3}{{8\,{\pi}}}\,\frac{{{\dot{M}}}}{{{\Omega}_{k}\,{\ell}^{3}}%
}\,\left(  1-\left(  \frac{r_{1}}{r}\right)  ^{1/2}\right)  . \label{eq12}%
\end{equation}

\section{Energy Conservation In The Disk}

The equation for the total energy conservation in the disk, as the matter
flows from $r_{2}$ to $r$, may be written as%

\begin{equation}
\frac{d}{dx_{k}}\left(  \rho\left(  \frac{1}{2}V_{\Phi}^{2}+\xi\right)
v_{k}-P\delta_{ik}v_{i}\right)  =-\rho\,v_{k}\,\frac{d\Psi}{dx_{k}}-\Lambda,
\label{eq13}%
\end{equation}
where $\xi$ is the total internal energy, $\Lambda$ is the amount of energy
per unit volume that leaves the system. Assuming the system is cooled by
radiation, we may write%

\begin{equation}
\Lambda=\frac{dF_{r}}{dz}. \label{eq14}%
\end{equation}

Inserting this expression into equation (\ref{eq13}) and using the equation of
continuity, we obtain%

\begin{equation}
\dot{M}\,\left(  \frac{1}{2}\,V_{\Psi}^{2}+\xi+\frac{{{\Omega}_{k}}^{2}%
\,{\ell}^{2}}{3}\right)  =\dot{M}\,\frac{G\,M}{r}-4\,\pi\,\int_{r_{2}}%
^{r}\,r\,F_{r}\,dr\,.\, \label{eq15}%
\end{equation}

Finally, assuming uniform viscous heat production along $z$, we write the
internal energy equation as%

\begin{equation}
\frac{d}{dt}\left(  {\rho}\,\xi\right)  =-P\,\frac{dv_{k}}{dx_{k}%
}-\frac{dF_{r}}{dz}+\frac{q^{+}}{{\ell}}\,. \label{eq16}%
\end{equation}

Now, making the approximation
\[
{\frac{dF_{r}}{dz}}\approx{\frac{F_{r}}{\ell}}\,,
\]
we solve for $F_{r}$ in equations (\ref{eq15}) and (\ref{eq16}), and equate
the results to obtain
\begin{equation}
-\ell V_{r}\,\frac{d}{dr}\left(  \rho\xi\right)  -\frac{P\,{\ell}}{r}%
\frac{d}{dr}\text{\thinspace}\left(  r\,V_{r}\right)  +q^{+}=-\frac{\dot{M}%
}{4\,{\pi}\,r}\,\frac{d}{dr}\,\left(  \frac{1}{2}V_{\Psi}^{2}+\xi
+\frac{{{\Omega}_{k}}^{2}\,{\ell}^{2}}{6}\right)  -\frac{{\dot{M}}}{4\,{\pi}%
}\,\frac{G\,M}{r^{3}}. \label{eq17}%
\end{equation}

This is the final energy equation for the disk, which has to be solved,
together with the hydrostatic equilibrium equation, to obtain the temperature
profile along the disk. It should be remarked that its deduction has been
achieved with no further assumption concerning the flow regime, i.e.,
supersonic or subsonic. We have, however, neglected terms of order of
$V_{r}^{2}$, and haven't considered transport of radiation along the radial direction.

\section{The Solution To The Energy Equation}

In the following we shall be solving equation (\ref{eq16}) with the main
concern to find out if there are geometrical thin solutions for the disk. It
is believed, under the $\alpha$-standard prescription, that far away from the
inner edge, in region dominated by gas pressure, the disk is geometrically
thin. Therefore, the following analysis is meant to this region. In other
words, we shall assume $r>>r_{1}$, and $P_{g}>>P_{r}$, which implies%

\[
\frac{{{\Omega}_{k}}^{2}{\ell}^{2}}{3}\approx\frac{k_{B}}{m_{H}}T\,,
\]
and
\[
\xi\approx\frac{\Omega_{k}^{2}\,\ell^{2}}{2}.
\]

Therefore, we may write%

\[
\rho\,\xi\approx\left(  \frac{3\,m_{H}}{2^{6}\,{\pi}^{2}\,k_{B}}\right)
^{1/2}\,\frac{{\dot{M}}\,{{\Omega}_{k}}^{2}}{{\alpha}\,T^{1/2}}\,,
\]%
\[
V_{r}\approx-\frac{k_{B}}{m_{H}}\,\alpha\,\frac{T}{r\,{{\Omega}_{k}}}\,,
\]%
\[
P\,\ell\approx\frac{1}{4\,{\pi}\,{\alpha}}\,\dot{M}\,\Omega_{k}\,,
\]%
\[
\ell\,V_{r}\approx-\sqrt{3}\left(  \frac{k_{B}}{m_{H}}\right)  ^{3/2}%
\alpha\frac{T^{3/2}}{r\,{{\Omega}_{k}}^{2}}\,,
\]
and
\[
r\,V_{r}\approx-\left(  \frac{k_{B}}{m_{H}}\right)  \alpha\frac{T}{{\Omega
}_{k}}\,,
\]
which, inserted into equation (17), yields%

\begin{equation}
\frac{dT}{dr}-\frac{14}{11}\frac{T}{r}=-\frac{8\,m_{H}}{11\,k_{B}}%
\frac{G\,M}{r^{2}}\,, \label{eq18}%
\end{equation}
the solution being%

\begin{equation}
T=A\,r^{{\frac{14}{11}}}+0.32\,\frac{m_{H}}{k_{B}}\frac{G\,M}{r}\,.
\label{eq19}%
\end{equation}

Using equations (15) and (19), we may write for the radiative flux,%

\begin{equation}
F_{r}=\frac{3}{40\,\pi}\frac{G\,M\,{\dot{M}}}{r^{3}}-\frac{70\,k_{B}}%
{88\,\pi\,m_{H}}\frac{\dot{M}}{r^{8/11}}\,A\,. \label{eq20}%
\end{equation}

Equation (\ref{eq20}) can be cast in a much more interesting form if we use
equation (\ref{eq19}) to express the constant $A$. We, then, obtain%

\[
F_{r}=\frac{\dot{M}}{8\,\pi\,r^{2}}\,\left(  \frac{29}{11}\,\frac{GM}%
{r}-\frac{70\,k_{B}}{11\,m_{H}}\,T\right)  \,,
\]
or%

\begin{equation}
F_{r}=\frac{\dot{M}}{8\,\pi\,r^{2}}\,\left(  \frac{29}{11}\,\frac{G\,M}%
{r}-\frac{70}{33}\,\Omega_{k}^{2}\,\ell^{2}\right)  .\, \label{eq21}%
\end{equation}

The first term in this equation is, practically, the same one that comes from
usual accretion disk theory, i.e.,
\[
\frac{3}{8\,\pi}\frac{G\,M\,\dot{M}}{r^{3}}\,.
\]
The second, that doesn't appear in the usual theory, displays the effect of
the temperature, or the height scale: the hotter, or the thicker, the less it
emits. It should be remarked that the largest temperature is
\[
T_{M}=\frac{29\,m_{H}}{70\,k_{B}}\,\frac{G\,M}{r}.
\]

In the standard formulation, one assumes equality between heat production and
cooling by radiation. Since the energy produced due to viscous dissipation,
$q^{+}$, is independent of the height scale, no matter how thin or thick the
disk may be, it always adjust itself to get rid of this energy. As compared to
this, the present formulation constitutes an improvement: it is clearly seen
that, if the disk thickens, it starts dimming.

\section{Radiation, Electronic Temperature and Height Scale}

If we write the energy exchange between protons and electrons as%
\begin{equation}
F_{ep}=\frac{3}{2}\frac{\rho k}{m_{H}}\frac{\left(  T_{i}-T_{e}\right)
}{t_{ep}}, \label{eq21a}%
\end{equation}
and equate it to the flux given by equation (\ref{eq21}), we obtain the time
protons and electrons need to reach thermal equilibrium,
\[
t_{ep}\approx\frac{3}{\alpha}\left(  1-\frac{T_{e}}{T_{i}}\right)  t_{dyn},
\]
$t_{dyn}$ being the dynamical time. We shall assume $\alpha<<1$, and
$t_{ep}>>t_{dyn}$.

The temperature we have obtained, due to our starting assumption, has to be
identified with the ion temperature. Clearly, this conclusion breaks down when
there is an abundant pair production and the pairs dominate the pressure. In
such a situation, it has to be identified with, what we improperly call, the
equivalent temperature, which we shall define at the right time. To know the
electronic temperature, we should assume a specific radiation mechanism.
Therefore, in the following, we shall make general considerations to link our
previous results to the radiative transfer problem. In our approach, we shall
try, to make a formulation not depending, as much as possible, on the values
of the accretion parameters.

Our starting point is to know the optical depth for thermal bremsstrahlung
absorption, $\tau_{ab}$. By definition \citep{fran92},%
\[
\tau_{ab}=6.22\times10^{22}\rho^{2}T_{e}^{-7/2}{\ell.}%
\]
We should expect $\tau_{ab}$ decreasing as we come close to the hole. Then, in
the inner region, the disk emits thin bremsstrahlung, and the flux is
\citep{ry79},
\begin{equation}
F=5\times10^{20}\rho^{2}T_{e}^{-1/2}{\ell}\text{.} \label{eq21b}%
\end{equation}
This yields%
\[
\tau_{ab}\approx1.38\times10^{-59}\frac{\dot{M}_{17}^{9}}{\alpha^{16}%
M_{10}^{11}}r^{5}%
\]
($r$ in units of $\frac{R_{g}}{2}$).

Finally, we obtain the electron scattering optical depth,%
\[
\tau_{sc}\approx\frac{0.225}{\alpha}\frac{\dot{M}_{17}}{M_{10}}r^{-1/2}.
\]

We conclude that if
\[
{\dot{M}}_{17}\leq3\,M_{10}\,,
\]
in such a way as to have the gas dominating the pressure, there may be three
regions in the disk, depending on the accretion parameters, $\alpha$, $M$,
$\dot{M}$:

a) The innermost region: $\tau_{sc}>\tau_{ab}$, and $\tau_{ab}<1$. In that
region, although $T_{e}<T_{i}$, it may be large enough to produce pairs. If
the pairs are very numerous, they may dominate the pressure. Then, we should
replace $T_{i}$ for the equivalent temperature, i.e.,%
\[
T_{eq}=T_{i}\left(  1+\left(  1+2y\right)  \frac{T_{e}}{T_{i}}\right)  ,
\]
where $y$ is the ratio pair density over ion density, that implies,%
\[
\Omega_{k}^{2}{\ell}^{2}=\frac{3k}{m_{h}}T_{eq}.
\]

To obtain the electronic temperature, we have to correct equation
(\ref{eq21b}) for the inclusion of pairs, i.e.,
\[
F=5\times10^{20}\rho^{2}T_{e}^{-1/2}{\ell}\left(  1+2y\right)  .
\]

This will lead to an equation relating $T_{e}$ and ${\ell}$, i.e.,%
\begin{equation}
\frac{A}{{\ell}^{10}}=\left(  \frac{29}{11}-\frac{70}{33}\left(  \frac{{\ell}%
}{R}\right)  ^{2}\right)  ^{2}, \label{eq21c}%
\end{equation}
where
\begin{equation}
A=5.13\times10^{41}\frac{\dot{M}^{2}}{\alpha^{4}\Omega_{k}^{6}}\left(
1+2y\right)  ^{2}T_{e}. \label{eq21d}%
\end{equation}

Clearly, for a given $T_{e}$, we may interpret it as an equation for $\ell$,
which will have solution as long as $A$ is less than a critical value $A_{c}$.
Solving under critical conditions, we obtain,%
\begin{equation}
6.95-11.185x^{2}+3.78x^{4}=0, \label{eq21e}%
\end{equation}
with $x=\frac{{\ell}}{R}$, the solutions being $x=0.942$ and $x=1.439.$The
solution $x=1.439$ leads to negative flux. Therefore, only $x=0.942$ is acceptable.

Equating $A$, given by equation (\ref{eq21d}), to $A_{c}$, given below,
\[
A_{c}=\left(  \frac{28}{33}\right)  ^{2}\Omega_{k}^{4}R^{14}x^{14},
\]
we obtain $T_{e}$. Therefore, in that region%
\[
\ell{\approx0.942}R.
\]

b) The intermediate region: $\tau_{sc}>\tau_{ab}>1$. In that case, the disk
will emit modified black-body-like, and the flux is \citep{ry79}
\begin{equation}
F^{MB}=2.3\times10^{7}\rho^{1/2}T_{e}^{9/4}. \label{eq21f}%
\end{equation}

Now the equation relating $T_{e}$ and $\ell$ will be%
\[
\frac{A}{{\ell}^{3}}=\left(  \frac{29}{11}-\frac{70}{33}\left(  \frac{{\ell}%
}{R}\right)  ^{2}\right)  ^{2},
\]
with
\[
A=7.98\times10^{2}\frac{T_{e}^{9/2}}{\alpha\Omega_{k}^{5}\dot{M}}.
\]
Proceeding in the same way, as we did before, we obtain%
\begin{equation}
6.95-11.185x^{2}-3.496x^{4}=0. \label{eq21g}%
\end{equation}
The solution now is $x=0.730$, the critical $A$ is%
\[
A_{c}=\left(  \frac{280}{99}\right)  ^{2}R^{3}x^{7},
\]
and%
\[
\ell{\approx0.730}R.
\]

c) The outer region: $\tau_{ab}>1$ and $\tau_{ab}>\tau_{sc}$. We shall assume
that $T_{i}=T_{e}$. If we assume the disk emits black-body-like, then%
\begin{equation}
-\frac{29}{11}+\frac{70}{33}x^{2}+Ax^{8}=0, \label{eq21h}%
\end{equation}
where%
\[
A\approx6.1\times10^{15}\frac{M_{10}^{2}}{\dot{M}_{17}r}.
\]
For $r=1000$, $M_{10}=\dot{M}_{17}=1,$%
\[
x\approx0.03.
\]

It should be remarked that in this region, there is no critical $A$, and
\[
\ell{\approx0.03}R.
\]

\section{Boundary Conditions and Flow Regime}

The expression for the radiative flux may be rewritten,%

\begin{equation}
F_{r}=\frac{\dot{M}\,V_{s}^{2}}{8\,\pi}\,\left(  \frac{29}{11}\,m_{\phi}%
^{2}-\frac{70}{33}\right)  \,, \label{eq22}%
\end{equation}
where $V_{s}$ and $m_{\phi}$ are, respectively, sound velocity and azimuthal
Mach number. Of course,
\[
m_{\phi}\geq0.9\,.
\]

Our problem, now, is to know how much the azimuthal Mach number can exceed the
above value. For that, let us assume that at the outer radius, $R_{d}$, the
temperature is $T_{d}$. Then,%

\begin{equation}
A=\left(  T_{d}-0.32\,\frac{m_{H}}{k_{B}}\frac{G\,M}{R_{d}}\right)
\,R_{d}^{-14/11}\,. \label{eq23}%
\end{equation}

If $\zeta$ is the efficiency for converting gravitational energy into radiation,%

\begin{equation}
\zeta\,\dot{M}\,c^{2}=0.3\,\frac{{\dot{M}}\,c^{2}}{r_{1}}-0.4\,\frac{k_{B}%
\,{\dot{M}}}{m_{H}}\,\left(  T_{d}-0.32\,\frac{m_{H}}{k_{B}}\,\frac{c^{2}%
}{r_{1}\,r_{d}}\right)  \,, \label{eq24}%
\end{equation}
or%

\begin{equation}
T_{d}=\frac{0.4\,m_{H}\,c^{2}}{k_{B}}\,\left(  \frac{1}{r_{1}}\,\left(
0.3-\frac{0.8}{r_{d}}\right)  -\zeta\right)  \,, \label{eq25}%
\end{equation}
$r_{1}$, and $r_{d}$ are, respectively, the inner and the outer radii in units
of half gravitational radius, ${\frac{R_{g}}{2}}$.

Using now the definition of the Mach number and the expression for the
temperature, we obtain%

\begin{equation}
m_{\phi}^{2}=1-\frac{3\,k_{B}}{m_{H}}\,\frac{R^{14/11}}{{V_{s}}^{2}}\,A\,,
\label{eq26}%
\end{equation}
which allows us to conclude that if $m_{\phi}>1$, $A<0$. Since, $T_{d}\geq0$,
we must have $\zeta\leq{\frac{0.3}{r_{1}}}$. If $\zeta$ is maximum,%

\begin{equation}
A=-0.32\,\frac{m_{H}\,G\,M}{k_{B}\,{R_{d}}^{25/11}}\,, \label{eq27}%
\end{equation}
which, inserted into equation (\ref{eq26}), leads to%

\begin{equation}
m_{\phi}^{2}=1+\left(  \frac{r_{d}}{r}-\left(  \frac{r}{r_{d}}\right)
^{14/11}\right)  ^{-1}\,\left(  \frac{r}{r_{d}}\right)  ^{14/11}\,.
\label{eq28}%
\end{equation}

Therefore, for maximum radiation efficiency, there will be a region in which
the flow regime is supersonic. However, even in that case, the disk will be
moderately supersonic in most of its extent. Just in a small outer fraction,
which depends on the size of the disk, will be really supersonic. It should be
remarked that maximum efficiency means null temperature (or almost) in the
outer border of the disk. To make things clear, let us calculate for a disk
with $r_{d}=200$, for: (a) $r=195$ and (b) $r=199$. We obtain:%

\begin{align*}
r  &  =195\rightarrow m_{\phi}=4.23\rightarrow\frac{\ell}{R}=0.236\,\\
r  &  =199\rightarrow m_{\phi}=9.40\rightarrow\frac{\ell}{R}=0.106\,.
\end{align*}

For different boundary conditions, such that $A>0$, the flow regime will be,
practically, sonic in the whole extent of the disk.

We have, here, a situation completely different from that we encounter in the
standard model. There, due to the equality between dissipation and cooling,
radiative flux obeys the same boundary condition imposed on the torque, and
has a maximum at $R=1.36 \, R_{1}$.

Inserting the value for $A$, obtained from the luminosity constraint, in the
expression for $T$, given by equation (\ref{eq19}), we obtain, at $r=r_{1}$,%

\[
T\approx5.84\times10^{11}\,,
\]
independent of the mass and accretion rate. For ${\beta}\approx1$, this
implies, for the azimuthal velocity, a Mach number of about 1. Therefore,
close to the inner radius,
\[
\frac{\ell}{r}\approx1\,.
\]

With these values,
\[
\frac{P_{r}}{P_{g}}\approx5\times10^{-18}\,\dot{M}\,,
\]
and our estimates hold as long as $\dot{M}\leq2\times10^{17}\,$g\thinspace
s$^{-1}$, for $M_{10}=1$. It is interesting to remark that the ratio of the
radiative flux to the heat flux due to viscous dissipation, close to the inner
radius, is%

\[
\frac{F_{r}}{q^{+}}\approx\frac{1}{5}\,.
\]
It should be remarked that, if some other cooling mechanism is considered
\citep{ml03}, this could be even less.

Analyzing the $r$ dependence of the radiative flux, we see that,
asymptotically, under the same boundary conditions,%

\[
\frac{F_{r}}{q^{+}}\approx1\,,
\]
the importance of radiative cooling grows as we go towards the outer radius.
Therefore, the geometrically thin disk assumption only holds in a very narrow
region where the flow is supersonic. As a matter of fact, the results we have
just obtained invalidates the geometrical thin disk assumption.

We should insist in a point: this region, where the flow is indeed supersonic,
only exists if radiation efficiency is maximum. In our formulation,
$\zeta=0.05$ and $\zeta=0.3$, respectively, for a Schwarzschild and a Kerr
hole. In that case,%

\begin{align*}
r  &  <0.88r_{d}\rightarrow m_{\phi}\leq2.0\,,\\
\,r  &  >0.88r_{d}\rightarrow m_{\phi}\geq2.0\,.
\end{align*}
If it is a little less, $A>0$ and the flow will be sonic, practically, all
over the disk.

Finally, it should be said that, since the radiative flux continuously grows
as we approach the hole, it can not satisfy the same boundary condition the
torque does. Therefore, no matter the correct boundary condition, the hole
does swallow, always, an amount of energy per unit time, which is, at least,
of the order of the disk luminosity.

\section{Conclusions}

We have used total and internal energy equations to obtain an equation for the
temperature along the disk. This equation solved under gas pressure dominance
shows a disk hotter than the standard one. Besides, it implies a radiative
flux decreasing with temperature or height scale. Though not considered here,
these findings impose severe constraints on the development of instabilities
in the disk. In that formulation, the role of advection is more important than
usually admitted. Independently of mass, accretion rate, and viscosity,
radiation transports, at the most, $20\,\%$ of the produced heat, close to the
inner radius. Far away, its efficiency may rise to $100\,\%$ . Under the
assumption of large radiation efficiency, as a boundary condition on the
solution for the temperature, it is shown that the flow is supersonic, the
azimuthal Mach number moderately exceeding 1in most of the extent of the disk.
Only, on a narrow outer region, of about $12\,\%$ of the disk extent, the Mach
number greatly exceeds 1. This highly supersonic regime is only obtained under
the assumption of TE or, at least, LTE, which is very unlikely to occur
\citep[and references therein]{hu00}. These results show that the
geometrical thin disk assumption fails everywhere in the disk, but in the
outer narrow region. However, even if the assumption of LTE holds in that
region, it represents a very small part of the disk.




\clearpage                                    

\end{document}